\documentclass[final]{article}
\usepackage[square,comma,numbers,sort&compress]{natbib}
\usepackage[T1]{fontenc}
\usepackage[latin1]{inputenc}
\usepackage[dvips]{color,epsfig,rotating}

\newcommand{\nsmall}{\fontsize{7}{9}\selectfont}

\begin{document}
\title{Spatio-Temporal Structuring of Brain Activity --- 
Description of Interictal EEG in Paediatric Frontal Lobe Epilepsy}
\author{W. Bunk$^{1*}$, T. Aschenbrenner$^{1}$, G. Kluger$^{2}$ and S. Springer$^{3}$} 
\date{}
\maketitle

1) Max-Planck-Institut für extraterrestrische Physik, Giessenbachstrasse,
85748 Garching, Germany. 
2) Klinik für Neuropädiatrie, Epilepsiezentrum, D-83569 Vogtareuth, Germany.
3) Heckscher-Klinikum für Kinder- und Jugendpsychiatrie und Psychosomatik,
München; Kinderklinik und Poliklinik im Dr. von Haunerschen Kinderspital, Klinikum der Universität München, Germany.\\
\mbox{{*}) Tel.: +49-89-30000-3548, Fax: +49-89-30000-3390, E-mail: whb@mpe.mpg.de}

\vspace{2cm}
\leftline{Running Title: Spatio-Temporal Structuring of Brain Activity}
\newpage %%%%%%%%%%%%%%%%%%%%%%%%%%%%%%%%%%%%%%%%%%%%%%%%

\begin{abstract}
\noindent A method for the quantitative assessment of spatio-temporal structuring
of brain activity is presented. This approach is employed in a longitudinal case study of a child with frontal lobe
epilepsy (FLE) and tested against an age-matched control group. 
Several correlation measures that are sensitive to linear and/or non-linear relations in multichannel
scalp EEG are combined with an hierarchical cluster algorithm.
Beside a quantitative description of the overall degree of synchronization the spatial relations are investigated 
by means of the cluster characteristics. The chosen information measures not only 
demonstrate their suitability in the characterization of the ictal and interictal phases 
but they also follow the course of delayed recovery of the psychiatric
symptomatology during successful medication. 
The results based on this single case study suggest testing this approach for
quantitative control of therapy in an extended clinical trial.

\noindent{\em Key terms: hierarchical cluster algorithm, information measures, 
  spatio-temporal structuring, synchronization, quantitative control of therapy}
\end{abstract}
\newpage %%%%%%%%%%%%%%%%%%%%%%%%%%%%%%%%%%%%%%%%%%%%%%%%

\section{Introduction}
In the last decade synchronization processes in the brain were put in the
focus of neurological science. They are claimed to be responsible for a variety of
essential properties of the brain: Synchronization is supposed to be a fundamental
method of information coding in the brain~\cite{synchronization_binding_and_expectancy} 
e.g. in visual perception. 
Obviously, in epileptic seizures synchronization plays an important role and
its quantitative description is relevant for diagnostic purposes~(\nocite{jerger_et_al0}%
\nocite{synchronization_and_information_flow_in_eegs_of_epileptic_patients}%
\nocite{clustering_approach_to_quantify_long-term_spatio-temporal_interactions_in_epileptic_intracranial_electroencephalography}%
\cite{jerger_et_al0}--\cite{clustering_approach_to_quantify_long-term_spatio-temporal_interactions_in_epileptic_intracranial_electroencephalography}). 
As the number of neurons
involved in this synchronized activity is rather high the associated surface EEG often
shows typical epileptiform patterns (spikes, waves) with considerable amplitudes. 
In interictal periods these typical epileptic EEG patterns are often missing which
does not contradict the diagnosis of epilepsy~\cite{pillai_sperling_0}.
Fingelkurts et al.~\cite{interictal_eeg_as_a_physiological_adaption_part1} 
hypothesize that chronic epilepsy changes the brains state even in interictal
periods displaying altered anatomical, biochemical and functional
properties. They claim that the interictal EEG, even without epileptiform
abnormalities, has a number of characteristic differences from the EEG of
healthy subjects and ``{\em a therapeutic intervention could aim to restore the actual composition of
brain oscillations and their temporal behavior, similar to those of the normal EEG\/}''~\cite{interictal_eeg_as_a_physiological_adaption_part1}.
Hence, an important aspect of interictal EEG assessment is
the evaluation of spatio-temporal synchronization. 
Up to now the analysis of synchronization is not viewed as a standard method for
diagnosing epilepsy based on interictal EEGs~\cite{pillai_sperling_0}.
We present in this paper a methodology to quantify 
synchronization properties of brain activity. A number of linear and non-linear
synchronization measures are employed to test for their suitability in this
question.

\noindent Frontal lobe epilepsy FLE is a common type of extra-temporal epilepsy 
in adults. In childhood, however, only less than 10\% of the developed (extra-temporal)
epilepsy syndromes are of this form~\cite{frontal_lobe_epilepsy_of_childhood_onset}.
Because of the considerable cognitive and behavioral deficits associated with FLE, 
anti-epileptic therapy is of greatest importance from the very beginning of the disease.\\
For a comprehensive diagnosis clinical signs, neuropsychological 
assessments, electroencephalographic findings, and magnetic resonance images have to 
be considered in a synoptic view. The anatomical assignment of clinical syndromes is controversial for FLE~\cite{salanova_etal0}.
Due to the variable and rapid spread of the epileptic activity the seizure characteristics can
vary strongly~\cite{fusco_iani_et_al0}. Seizures with origin in the frontobasal cingulate area
often include complex motor automatisms~\cite{salanova_etal0}. Common and striking clinical
features related to FLE especially in children are atypical absences and psychomotor seizures,
characterized by complex (organized) movements of the limbs and an increased general motor
activity~(\cite{luders_burgess_noachtar},\cite{psychomotor_seizures_of_temporal_lobe_onset}). Seizures originating from the frontal
lobe are usually of short duration, frequent, and often arise during sleep~\cite{luders_burgess_noachtar}.  
The onset of the FLE in childhood is very frequently misdiagnosed 
as sleep disorder or as a psychogenic seizure~(\cite{fusco_iani_et_al0},\cite{fohlen_bulteau0},\cite{kellinghaus_lueders0}).
Neuropsychological testing and assessments
regarding behavioral deficits such as increasing learning disability and forgetfulness
give further indications on the status during the ongoing 
disease~(\nocite{unilateral_frontal_lobe_epilepsy_affects_executive_functions_in_children}\nocite{elger_brockhaus0}\nocite{harvey_et_al0}%
\cite{unilateral_frontal_lobe_epilepsy_affects_executive_functions_in_children}--\cite{harvey_et_al0}). The observed behavioral disorders
are explained by the frequent subclinical epileptic activity in the prefrontal area visually
detectable only with intracranial EEG recordings~\cite{fohlen_bulteau0}.\\
The ictal electroencephalographic monitoring of FLE patients yields typical features of 
epileptic discharges in general, e.g. low-amplitude-fast-activity, rhythmic spikes and waves~\cite{fohlen_bulteau0}. 
Particularly, one observes ictal and pre-ictal synchronization 
effects over wide areas of the brain. In the interictal phase the scalp-EEG 
is mostly normal, i.e. by means of a conventional visual inspection 
suspicious features can hardly be found. Even from the ictal EEG, it is often difficult
to distinguish between FLE and temporal epilepsy~\cite{manford_fish_shorvon0}. For a more precise localization
of the epileptic focus an ictal and intracranial recording is needed.   
Because of the rapid spread of the pathological patterns 
the origin of the epileptic discharge is hard to detect even with invasive techniques.\\
The objective of this paper is to present a methodological
approach to the analysis of scalp electrode EEG for a characterization of 
spatio-temporal structuring of brain activity.\\
The paper is organized as follows:\\
Section 2 gives a description of the data together with a clinical
 history of the patients and some technical aspects of the 
electroencephalographic data acquisition. In section 3 follows an detailed explanation 
of our methodological approach for the study of brain synchronization. It comprises the definition 
of the various correlation measures used in frequency and time domain, 
the derived similarity measures and the hierarchical clustering approach. 
Section 4 presents our quantitative results and introduces a visualization of the cluster behavior 
by a two-dimensional projection onto the head surface.
The paper ends with a short summary of the results. Main emphasis is placed in the
discussion on the observed specific spatio-temporal structuring in the patient with FLE.

\section{Patients and Data}
\subsection{Patients}
In this study the single case results of the analysis of several EEG recordings covering
two years of a child (patient $A$) are compared with the outcome of an age-matched control group.\\
Patient $A$ has a family history of epilepsy. Until the age of
eleven years the boy's motor and cognitive development was in the normal range. The paediatric and neurological
investigation never showed pathological results.  At 11.5 years the boy
suffered from atypical absences with enuresis but without spasms. 
The absences were followed by complex partial seizures (11.6 years). At the age of
11.8 years frequent typical hypermotoric seizures appeared. Since this time
his psychopathological status deteriorated rapidly: The boy became disorientated
and aggressive. Other clinical features of his aggravated status were speech disorders,
extremely shortened attention span, reduced emotional control, motor hyperactivity
and depressive mood. These clinical manifestations and the findings from the
EEG recordings (spike waves, low-amplitude-fast-activity during seizures) are consistent with the diagnosis of a frontal lobe epilepsy
generated in the frontobasal area. MRI of the brain, laboratory diagnostic of serum, and CSF were inconspicuous. \\ 
Regular medication started at the age of 12.1 years with oxcarbazepine (15\ mg/kg/d), which induced a first status
epilepticus. This medication was immediately stopped and replaced by high dose
i.v.-valproic acid and i.v.-phenobarbital. At
12.15 years it was possible to  obtain seizure freedom with  valproic acid
(22\ mg/kg/d) and lamotrigine (5\ mg/kg/d). Half a year after the beginning of the medication (12.7 years) 
reduced cognitive capability with learning disability (IQ 85) was measured
using Raven Test.\\ 
As a consequence of dose reduction atypical absences occurred
again at the age of 12.5 to 12.9 years (not documented by EEG) which required increased
lamotrigine doses. The boy is seizure free since the age of 14.0 years with an ongoing
lamotrigine monotherapy.\\
Patient $A$ had an unsuspicious occipital background activity within the
alpha-band. This background activity, assessed by visual EEG inspection
in interictal phases, did not change significantly during the disease. 
The first four EEG, which entered this study, are taken at an age of 12.1
years and cover the first month of medical therapy (acute phase). 
In the first EEG a seizure was derived. Pre-ictal, a short episode of generalized low-amplitude-fast-activity was
recorded. During seizure (one minute) rhytmic spike wave complexes were seen as
most prominent ictal patterns. In the course of the second EEG recording
another seizure episode occured. Starting at $t\approx 6.7\ min$ (compare figure~\ref{cor_plot_single_trace_ami}),  
epileptic activity persisted with rhythmic generalized delta activity most pronounced in frontal areas. The EEG seizure patterns were similar to the first attack.
During the following six EEG recordings no seizure took place. Until the age of 12.1 years
typical interictal bifrontal epileptiform patterns (like spikes and spike
wave complexes) were detected visually. The third
and the fourth EEG showed only sporadic atypical irregular sharp waves. The subsequent four EEG were obtained at 
an age of 13.15, 13.35, 13.73 and 13.96 years. 
These four EEG recordings (including the one during clinical deterioration at 13.15 years) showed no epileptiform patterns.
The EEG recordings span a period of almost two years of successful therapy. \\
The age-matched control group consists of three patients whose EEG recordings cover an age
range from 10.6 to 14.4 years.  
Patient $B_1$ of this group was diagnosed with an uncomplicated generalized
epilepsy. He suffered from three unprovoked generalized seizures before
starting treatment and is seizure-free with a monotherapy of valproic 
acid. The patients $B_2$ and $B_3$ experienced only one unprovoked generalized epileptic seizure. 
Ten EEG were taken from patient $B_1$ (10.6 to 13.8 years) and one EEG
from $B_2$ (11.3 years) and $B_3$ (14.4 years), respectively. All
EEG recordings of the control group were classified as normal. 
These three patients showed seizure freedom for more than five years
of follow-up. In contrast to patient $A$ their cognitive and emotional
development was in the normal range. In this sense it was an unbiased selection
of controls matching the age of patient $A$.\\

\subsection{Data}
The positioning of the electrodes followed that of the
standardized 10-20--Inter\-national System of Electrode Placements. 
Every EEG recording consists of 21 synchronously obtained time series. Our
data base is made up by a number of twenty multichannel EEG recordings: Eight EEG are recorded
from patient $A$ (age range 12.06--13.96 y). Twelve EEG are derived from the control group (three patients,
age range 10.61--14.37 y). Every EEG record measures brain activity for at
least 10 minutes at a sampling rate of 250~Hz and a signal depth of 16~bits.
Information about patients and EEG recordings are listed in table~\ref{patients_data}.\\
To guarantee the comparability, the patients state during EEG monitoring\label{data_preprocessing}
was classified with a time resolution of one second by an expert. 
Components of the patients state assessed were e.g. vigilance, attention
level, artefacts (movements) and epileptic activity. This expert was  not directly involved in the subsequent
analysis. The selected EEG-segments (free of artefacts and movements, awake,
same level of attention, no interictal epileptiform discharges or ictal EEG patterns) form a homogenous data base which enters the study.\\

\begin{sidewaystable}
\arrayrulewidth1.0pt
\renewcommand{\arraystretch}{1.1}
{\nsmall 
\begin{tabular}{|c|c|r|c|c|c|r|l|l|l|}
\hline
\multicolumn{2}{|c|}{Patient/\hphantom{Gr}} & \multicolumn{1}{|c|}{$\Delta $T}& Age & State of & Seizure& \multicolumn{1}{|c|}{Seizure} & \multicolumn{1}{|c|}{Medication} & \multicolumn{1}{|c|}{Clinical Diagnosis} & \multicolumn{1}{|c|}{EEG Findings} \\ 
\multicolumn{2}{|c|}{\hphantom{Pa}Group} &  \multicolumn{1}{|c|}{[s]} &  [y] & Vigilance & during EEG&\multicolumn{1}{|c|}{Freedom [h]}&&&\\ 
\hline
\hline
             & SC/acute phase & 710/125 & 12.060 & AW/SL & $+$ &$<1$& OXC & FLE -- mult. seizures & SW frontal + seizure patterns\\
\cline{2-10}
             & SC/acute phase & 940/340 & 12.096 & AW & $+$ &$<1$& VPA + LTG & FLE -- seizures & SW frontal + seizure patterns\\ 
\cline{2-10}
             & SC/acute phase & 3600/769 & 12.104 & AW/SL & $-$ &$>24$& VPA + LTG & FLE -- only atyp. absences & sporadic SW\\ 
\cline{2-10}
             & SC/acute phase & 990/932 & 12.159 & AW & $-$ &$>100$& VPA + LTG & FLE -- seizure free & sporadic SW\\
\cline{2-10}
\raisebox{2.0ex}[-2.0ex]{A} & SC & 900/435 & 13.151 & AW & $-$ &$>100$& VPA + LTG & FLE -- seizure free & inconspicuous\\
\cline{2-10}
             & SC & 970/921 & 13.351 & AW & $-$ &$>1800$& VPA + LTG & FLE -- seizure free & inconspicuous\\ 
\cline{2-10}
             & SC & 930/615 & 13.729 & AW & $-$ &$>5100$& VPA + LTG & FLE -- seizure free & inconspicuous\\ 
\cline{2-10}
             & SC & 980/648 & 13.964 & AW & $-$ &$>7200$& LTG & FLE -- seizure free & inconspicuous\\ 
\hline
\hline
             & CG & 890/761 & 10.608 & AW & $-$ &$>24$& NoMed & after 1. unprovS & inconspicuous\\ 
\cline{2-10}
             & CG & 1900/588 & 10.877 & AW/SL & $-$ &$>24$& NoMed & after 2. unprovS & inconspicuous\\
\cline{2-10}
             & CG & 660/583 & 10.918 & AW & $-$ &$>380$& NoMed & after 2. unprovS & inconspicuous\\ 
\cline{2-10}
             & CG & 1680/544 & 11.301 & AW/SL & $-$ &$>3700$& NoMed & after 2. unprovS & inconspicuous\\
\cline{2-10}
             & CG & 900/782 & 11.323 & AW & $-$ &$>3900$& NoMed & after 2. unprovS & inconspicuous\\ 
\cline{2-10}
\raisebox{2.0ex}[-2.0ex]{B$_1$}  & CG & 1030/785 & 11.858 & AW/SL & $-$ &$>100$& NoMed & after 3. unprovS & inconspicuous\\ 
\cline{2-10}
             & CG & 1030/856 & 11.929 & AW/SL & $-$ &$>700$& VPA & GE -- seizure free & inconspicuous\\ 
\cline{2-10}
             & CG & 960/944 & 12.299 & AW & $-$ &$>3900$& VPA & GE -- seizure free &inconspicuous\\ 
\cline{2-10}
             & CG & 900/824 & 12.792 & AW & $-$ &$>8200$& VPA & GE -- seizure free &inconspicuous\\ 
\cline{2-10}
             & CG & 880/755 & 13.751 & AW & $-$ &$>16600$& VPA & GE -- seizure free &inconspicuous\\
\hline
B$_2$  & CG & 900/879 & 11.282 & AW & $-$ &$>24$& NoMed & after 1. unprovS & inconspicuous\\ 
\hline
B$_3$  & CG & 930/841 & 14.373 & AW & $-$ &$>24$& NoMed & after 1. unprovS & inconspicuous\\ 
\hline                                     
\end{tabular}   
}
\caption{\small\em Description of patients and data. Abbreviations: SC: Single Case, CG: Control Group; 
$\Delta $T: Length of recording/duration of selected EEG-segments according to
the definitition in section~\ref{data_preprocessing}; AW: awake, SL: sleep (periods during sleep were skipped in the analysis); 
NoMed: no medication, OXC: oxcarbazepine, VPA: valproic acid, LTG:
lamotrigine; FLE: frontal lobe epilepsy, GE: generalized epilepsy, unprovS:
unprovoked seizure; SW: sharp waves.
\label{patients_data} 
}
\end{sidewaystable}

\section{Methods}
To obtain a time dependent characterization of brain activity the data has been 
analyzed with a sliding window technique. Since the window size $n_0$ at a
given sampling rate determines the frequency bandwidth we chose $n_0=512$, corresponding to a real time period of slightly more 
than two seconds. The consecutive windows overlap by half the window size, which results
in more than 900 separate windows for a typical EEG recording of $15$ minutes 
and a time resolution of $\sim$1.0 seconds.\\
To describe the spatio-temporal cluster properties, at each instant of time $t$ (center of window) 
several correlation measures between data of the several leads were
computed. From the respective quantity appropriate distance measures were derived. Finally, 
the resulting distance matrices were  evaluated in a cluster analysis in
analogy to Bonanno et al.~\cite{bonlilman0} (later adopted in the context of
EEG-analysis by Aschenbrenner et al.~\cite{eeg_2005} and Lee et al.~\cite{classification_of_epilepsy_types_through_global_network_analysis_of_scalp_electroencephalograms}).\\
Details of this procedure are explained in the following subsections. 

\subsection{Correlation and Distance Measures}
The correlation between each pair of time series was estimated in several
ways. Hereby, we distinguish between measures in the frequency domain (using Fourier
transform) and time domain when applying correlation algorithms directly to the
time series. In the latter case we particularly studied the differences between 
linear and non-linear correlation measures.

\subsubsection{Correlation Measures in the Frequency Domain}
Fourier transform is applied to succeeding data segments ($n_0$ data points),
generated by the sliding window procedure and  using window sampling techniques
(Hamming windowing). For further analysis the normalized power
spectra $F^{X}_{\;t\;}(\nu _{i})$ of each window of time series $X$ were 
taken, considering only the spectral range between $0 \le \nu _i \le 30\ Hz$, 
so that $\sum_{\nu _i=0}^{30Hz}F^{X}_{\;t\;}(\nu _{i})=1$.\\ 
The ``correlation'' between spectra of $X$ and $Y$ was then estimated by means of the 
{\em Kullback-Leibler entropy}~\cite{kullback_leibler_0}: 
\begin{equation}
E^{KL}_{\;t\;}\left(X,Y\right) =\sum_{\nu _i=0}^{30Hz} F^{X}_{\;t\;}(\nu _{i})\log_{2}\left(\frac{F^{X}_{\;t\;}(\nu _{i})}{F^{Y}_{\;t\;}(\nu _{i})}\right)  
\label{KullbackLeibler}
\end{equation}
The Kullback-Leibler entropy measures the dissimilarity between two
statistical distributions based on a generalization of Shannon's definition of information entropy.
Since $E^{KL}_{\;t\;}\left(X,Y\right)$ is usually not symmetric, we symmetrized 
$E^{KL}_{\;t\;}\left(X,Y\right)$ according to the ``resistor-average distance''~\cite{symmetrizing_kullback_leibler_0}:\newline
\begin{equation}
d\left(E_{t}\left(X,Y\right)\right)=d\left(E_{t}\left(Y,X\right)\right)=\left(\frac{1}{E^{KL}_{\;t\;}\left(X,Y\right) }+\frac{1}{E^{KL}_{\;t\;}\left(Y,X\right) }\right) ^{-1}
\label{symm_KullbackLeibler}
\end{equation}
In this modified form, the symmetrized Kullback-Leibler entropy can serve as a kind of a
distance measure: $d\left(E_{t}\left(X,Y\right)\right)=0$ iff 
$F^{X}(\nu _{i})=F^{Y}(\nu _{i}) \forall \nu _{i}$, it is $> 0$ else. This
measure emphasizes similarities in the spectral content of both time series
while phase aspects are ignored.\\
%%%%%%%%%%%%%%%%%%%%%%%%%%%%%%%%%%%%%%%%%%%%%%%%%%%%%%%%%%%%%%%%%%%%%%%%%%%%%%%%%%%%%%%%%%%%%%%%%%%%%
%%%%%%%%%%%%%%%%%%%%%%%%%%%%%%%%%%%%%%%%%%%%%%%%%%%%%%%%%%%%%%%%%%%%%%%%%%%%%%%%%%%%%%%%%%%%%%%%%%%%%

\subsubsection{Correlation Measures in the Time Domain}
To quantify synchronization between pairs of time series $X$ and
$Y$ with respect to their temporal behavior some correlation measures are tested.
The most common measure is {\em Pearson's coefficient of correlation}:
\begin{equation}
\rho_{t}\left(X,Y\right) =\frac{\sum^{n_0-1}_{i=0} (X_{i}-\overline{X})(Y_{i}-\overline{Y})}
 {\sqrt{\sum^{n_0-1}_{i=0} (X_{i}-\overline{X})^{2}\sum^{n_0-1}_{i=0} (Y_{i}-\overline{Y})^{2}}}
\label{pearsons_corr}
\end{equation}
Even so $\rho$ is independent of the origin and the
scale it assumes a mutual linear correlation between $X$ and $Y$, and 
normality of the distributions.
In order to overcome this restrictions, the {\em Spearman rank-order correlation coefficient}
can be used: 
\begin{equation}
\rho^{\prime }_{t}\left(X,Y\right)=1-6 \frac{srd}{n_0(n_0^2-1)}. 
\label{spearmans_corr}
\end{equation}
In this expression $srd$ denotes the sum of the squared difference in
the rank ordering of the corresponding variables $X$ and $Y$. This distribution-free correlation measure 
is not restricted to linear correlations and is robust with
respect to the outlier-problems in time series.\\

\noindent We focus, however, on the use of an information based
correlation measure, the mutual information $M$. Its definition
is closely related to Shannon's information entropy $H$.

\begin{eqnarray}
M_{t}\left(X,Y\right)&=&\sum_{i=0}^{nb-1}\sum_{j=0}^{nb-1}p_{t}(X_{i}Y_{j})\log_{2}{\frac{p_{t}(X_{i}Y_{j})}{p_{t}(X_{i})p_{t}(Y_{j})}}\\
&=&H_{t}\left(X\right)+H_{t}\left(Y\right)-H_{t}\left(X,Y\right)
\label{mutual_info}
\end{eqnarray}
Herein, $H_{t}\left(X\right)=-\sum_{i=0}^{nb-1}p_{t}(X_{i})\ log_{2}\;p_{t}(X_{i})$ is the 
so-called marginal entropy, i.e. Shannon's information entropy applied to $X$. The mixed term $H_{t}\left(X,Y\right)$ in 
Eq.~\ref{mutual_info} is called joint entropy. 
$nb$ gives the number of discrete bins of the corresponding distribution functions $p_{t}(X)$ and
$p_{t}(Y)$. The discretization of $X$ and $Y$ has not to be homogeneous. 

\noindent If we calculate the mutual information for every pair of EEG recordings we end up with a
symmetric mutual information matrix $\widehat{M}$.
Major differences to the correlation coefficients of Pearson and Spearman are that 
the mutual information does not distinguish between positive and negative correlations and
that it is less sensitive to phase shifts between the data sets. 

\subsection{Similarity and Dissimilarity Measures}
\label{SimMeasures}As similarity measures (e.g. correlation coefficient, mutual information)
are never metric~\cite{legendre_legendre0} %[p.253]
for clustering purposes these measures have to be transformed to dissimilarity metric
measures (``distances''). All measures of correlation $Q$ can be transformed to
an appropriate distance matrix $\widehat{D}$ with elements 
$d\left(Q_{t}\left(X,Y\right)\right)$ that allows for the investigation of
similarity clustering. In the case of Pearson's coefficient, we follow the
proposal of Bonanno et al.~\cite{bonlilman0}. %[p.97]
The distance measure for each epoch $t$ is given by 
\begin{equation}  
d\left(\rho_{t}\left(X,Y\right)\right)=\sqrt{2\left(1-\rho_{t}\right)}.
\label{pearson_spearman_dist}
\end{equation}
$d(\rho)$ takes values between $0$ (perfectly correlated, $\rho=1$) and $\sqrt{2}$ 
(perfectly anti-correlated, $\rho=-1$).
\\
For Spearman's coefficient, the distance measure is given by the equivalent
expression. $d\left(\rho\right)$ and $d\left(\rho^{\prime}\right)$ fulfill the 
requirements of a metric% 
%%%%%%%%%%%%%%%%%%%%%%%%%%%%%%%%%%%%%%%%%%%%%%%%%%%%%%%%%%%%%%%%%%%%%%%%%%%%%%%%%%%%%
\footnote{A metric is a symmetric, positive quantity --- minimal value $0$ achieved
for identical elements ---, which meets the triangle
inequality~\cite{legendre_legendre0}. %[p.274]
}%
%%%%%%%%%%%%%%%%%%%%%%%%%%%%%%%%%%%%%%%%%%%%%%%%%%%%%%%%%%%%%%%%%%%%%%%%%%%%%%%%%%%%%
.

\noindent In the case of the mutual information we use as a distance measure 
\begin{equation}  
d\left(M_{t}\left(X,Y\right)\right)=\log_{2}nb-M_{t}\left(X,Y\right).
\label{mutualdist}
\end{equation}

\noindent The Kullback-Leibler entropy is a dissimilarity (distance-like) measure by definition 
and therefore a further transformation is not necessary. 
Even the symmetrized version of the Kullback-Leibler entropy (see Eq.~\ref{symm_KullbackLeibler}) 
is non-metric in a rigid sense as it still violates the triangle inequality. However, 
the application of the Kullback-Leibler entropy in the framework of section~\ref{temp_char} does not
require metricity.

\subsection{Hierarchical Clustering}
\label{SectHierarch} For the description of the overall relationship 
between the EEG measurements at the different locations in form of an 
indexed tree we apply a single linkage cluster analysis SLCA~\cite{GowerandRoss1969}.   
The definition of the respective hierarchy $H = \{h_0,h_{1},...\}$ is based 
on the interobject dissimilarities derived from the correlation measures 
as shown in the previous section. 

\noindent Given a distance measure $d\left(Q_{t}\left(X,Y\right)\right)$ 
(e.g. in form of a distance matrix) the reconstruction of a hierarchical tree 
in the sense of a SLCA approach is straight forward with the help of the
so-called minimum spanning tree (MST)~\cite{GowerandRoss1969}.
One possible way of the reconstruction of a hierarchical tree is 
briefly summarized here (for details see~\cite{GowerandRoss1969} or~\cite{ramtouvir0}):\\
In the first step of generating the hierarchy, the minimum spanning tree has
to be constructed. Remember, for a number $n$ of objects the MST is the shortest connection
of all these objects without any loops and it consists of $n-1$ edges $e_{ij}$.
(Most algorithms for generating the MST are based on the concepts of Kruskal~\cite{kruskal_56} or Prim~\cite{prim_57}.) 
The resulting tree is represented by a sequence of edges  $\left\{e_{ik},...,e_{lj}\right\} $. 
Hereby, the length of edge $e_{ik}$ is equivalent to the %``true''
distance $d_{XY}$ as introduced in the previous section. 
The ultrametric distance $d_{XY}^{\ast}$ between two arbitrary
elements $X$ and $Y$ is then given by the longest edge of the sequence
$\left\{ e_{Xk},...,e_{lY}\right\} $ that connects the vertex $X$ with the
vertex $Y$ along the tree. 

\noindent The process of generating a hierarchical tree can be understood as
an iterative splitting process, starting at the highest aggregation level at
which all elements belong to the same hierarchy $h_{0}$. 
At this level $f(h_{0})$ is the maximum value of $d^{\ast}\left (Q_{t}\left (X,Y\right )\right )$. 
At the next (lower) level the set of elements is divided into 
these two sets that are separated by the longest ultrametric distance. 
By construction, this is a realization of the SLCA-approach, guaranteeing
that the distance between two clusters equals the minimum intercluster distance
between arbitrary elements of these clusters.
 This iterative procedure is repeated with ever
decreasing thresholds (ultrametric distances) of the generated subtrees 
from top to bottom until $f(h_{k})=0$, indicating that the resulting hierarchy consist only of one element.

\noindent We want to emphasize that other approaches than the SCLA method 
could be chosen for the agglomerative algorithm, 
e.g. by applying  complete linkage distances or the
use of the distances between cluster centroids.  In this context, however, 
we explicitly decided for this approach because it is adequate for
the description of the dynamical propagation of brain activities between 
distant areas.
    
\noindent The left panel in figure~\ref{patient_A_before_seizure},
\ref{patient_A_during_seizure} and~\ref{Control_group_B1_6}, respectively, shows a visualization of 
the corresponding hierarchical tree. 
The numerical aggregation levels, equivalent to the defined cluster 
distances,  are drawn at the ordinate.

\section{Results}
Applying the described techniques to EEG recordings, we obtain a time-resolved information
on spatial characteristics of electrical brain activity. In a first step, however, the analysis considers
the overall level of synchronization --- ignoring time-dependence and local information --- in the 
different patients and EEG recordings. We investigated whether such global measures are able to distinguish between the single
case and the control group even in long lasting interictal periods. Finally, we focus on the results from the clustering procedure, 
which yields spatio-temporal characteristics of brain activity. Locally resolved properties of
the processes provide further details to characterize brain activity and 
furthermore could give some hints on the localization of pathologies.  

\subsection{Temporal Characterization}\label{temp_char}
\begin{figure}[htb]
\epsfclipon
\epsfxsize=12.51cm
\centerline{\epsffile[51 258 546 570]{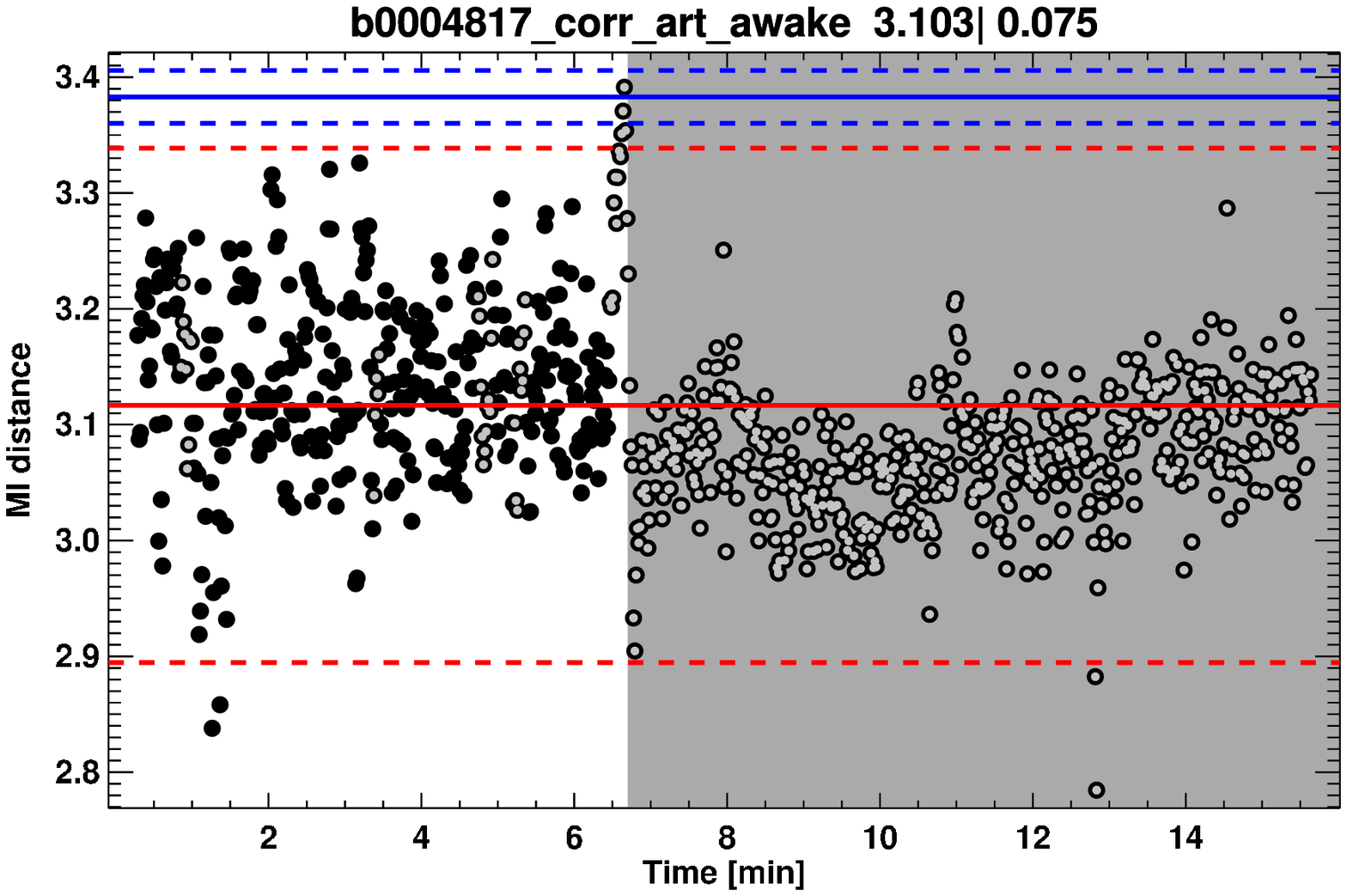}}
\caption{\small\em 
  Time course of the mean distance $D\left(M_{t}\right)$ --- based on the
  mutual information --- for the second EEG recording of patient A
  (age$=12.10\ yr$). The according value of the averaged mean distance 
  $D\left(M_{t}\right)=3.103$ and standard deviation 
  $\sigma_D\left(M_{t}\right)=0.075$ (compare second observation of patient $A$ in
  figure~\protect\ref{mutual_mean_age}). At about $6.7\ min$ an epileptic 
  seizure (hatched area) starts. 
  The seizure attack is initiated by a loss of synchronization,
  while during the seizure the correlation is generally increased, thus the
  distance values are low. The red horizontal lines mark the mean value and standard
  deviation of the different EEG recordings of patient A. The blue horizontal
  lines label the corresponding quantities for the control group. Periods
  with interictal epileptiform discharges or ictal patterns, labeled in gray, are eliminated and do not enter
  the statistics.
\label{cor_plot_single_trace_ami}
} 
\end{figure}

For every epoch (window) labeled with index $t$ we calculated the distance matrix
$d_t$ based on  the respective similarity measure $Q$ as defined in
section~\ref{SimMeasures}. The mean of all entries of the distance
matrix for each epoch $t$ yields a new measure $\overline{d\left(Q_{t}\right)}$ that 
quantifies the general degree of synchronization at each time $t$, whilst the
spatial information is not yet considered here explicitly. Figure~\ref{cor_plot_single_trace_ami} displays the time 
course of $\overline{d\left(M_{t}\right)}$, i.e. the distance measure using the mutual 
information, for the second record of patient A. Note, that in the segment prior to
seizure the distant values are significantly lower than in the control group.
At $t\approx 6.7\ min$ starts a psychomotor seizure, which is reflected by a
significantly decreased value of $\overline{d\left(M_{t}\right)}$: Apparently, the 
transition to seizure (marked by the gray background) coincides with a sharp rise and drop in the distance 
measure, indicating a short-time desynchronization followed by a persistent higher degree of synchronization.\\ 
In order to characterize the degree of synchronization between all 21
lead-areas of the multi-channel EEG for each patient and each record, we 
average over the spatial mean: 
${D\left(Q\right)}=\frac{1}{T} \sum_{i=1}^{N}\overline{d\left(Q_{i}\left(X,Y\right)\right)}$. The according 
standard deviation is abbreviated by $\sigma_D\left(Q\right)$.
We observe significant differences between patient $A$ (acute phase) and the control group in 
the sense that the correlation distances are
smaller in general. However, this effect is most distinctive by using either
the spectral description in the frequency domain combined with the non-linear
similarity measures (according to  Eq.~\ref{symm_KullbackLeibler}) or by use of
the non-linear correlation measure, namely the mutual information (see
Eqs.~\ref{mutual_info} and \ref{mutualdist}). 

\begin{figure}[htb]
\epsfclipon
\epsfxsize=12.51cm
\centerline{\epsffile[51 258 546 569]{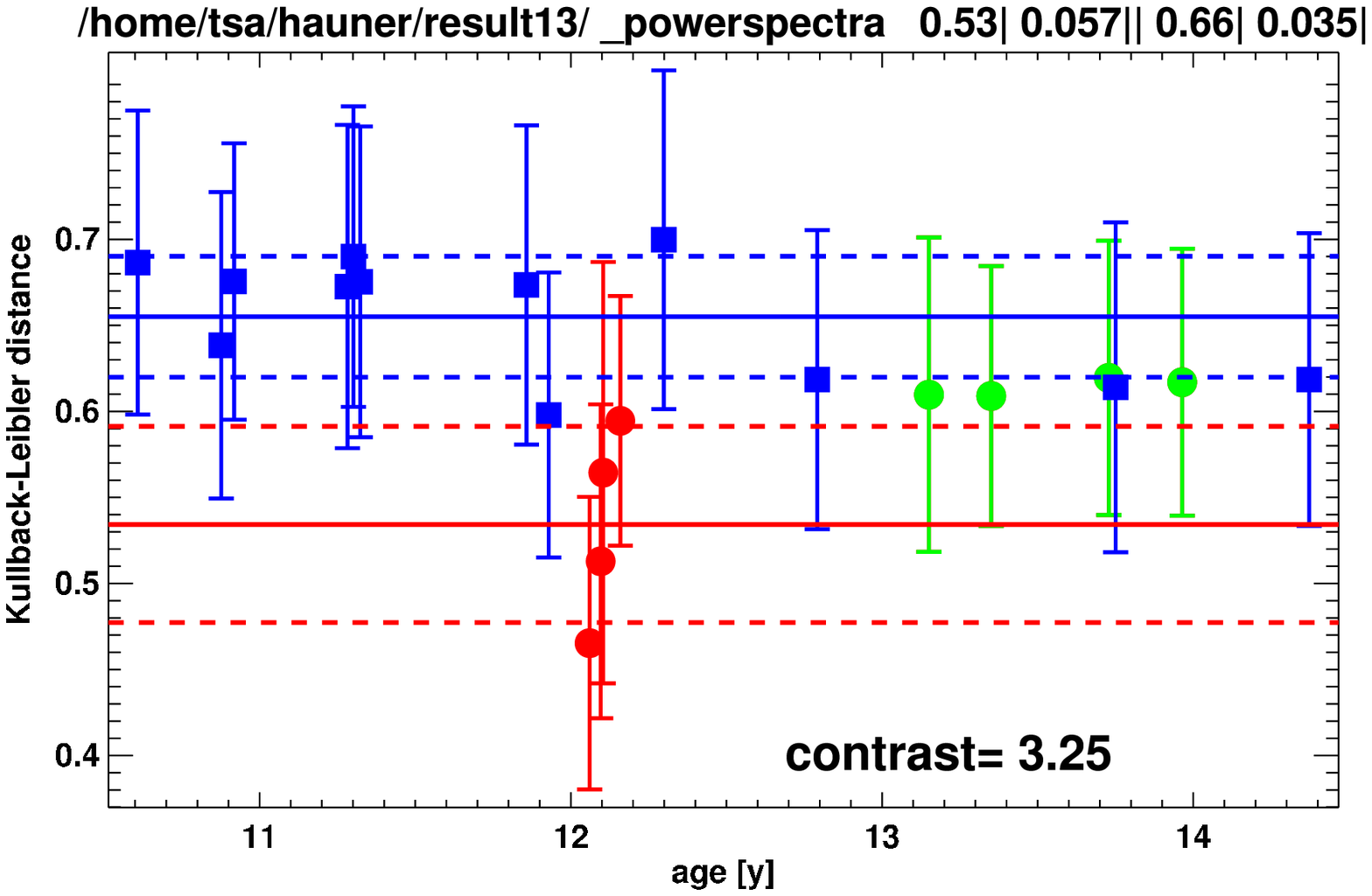}}
\caption{\small\em Mean Kullback-Leibler distance for patient $A$ (acute phase: red,
  non-acute phase: green) and control group (blue) of the power spectra. The filled circles denote the averaged mean
  Kullback-Leibler distance $D\left(E\right)$, the error bars indicate the range 
  $\pm \sigma_D\left(E\right)$ of the single
  EEG recordings. The horizontal lines label the mean Kullback-Leibler
  distance for patient $A$ and control group, the dashed lines label the respective
  standard deviations.   
\label{power_mean_age} 
}
\end{figure}

\begin{figure}[htb]
\epsfclipon
\epsfxsize=12.51cm
\centerline{\epsffile[51 258 546 569]{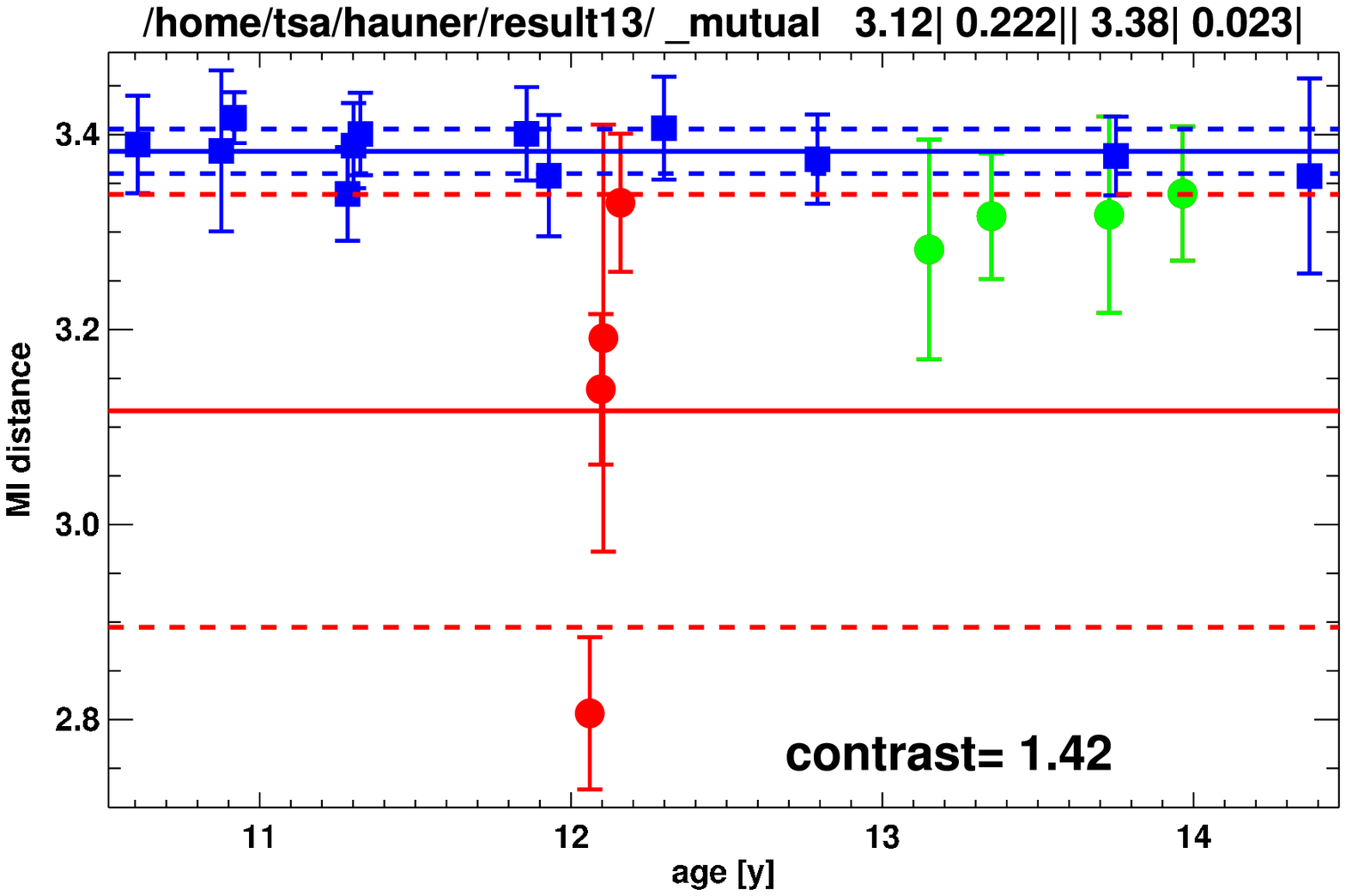}}
\caption{\small\em Mean mutual Information for patient $A$ (acute phase: red,
  non-acute phase: green) and control group
  (blue). The filled circles denote the averaged mean mutual information
  $D\left(M\right)$, the error bars indicate the standard deviations $\sigma_D\left(M\right)$
  of the single EEG recordings. The horizontal lines label the mean mutual
  information for patient $A$ and control group, the dashed lines mark the
  respective standard deviations.   
\label{mutual_mean_age} 
}
\end{figure}

\noindent Moreover, the diagrams in figure~\ref{power_mean_age} and \ref{mutual_mean_age}
show that these obvious differences vanish during
treatment and with time quite monotonically. This trend is obvious in both non-linear measures, using Kullback-Leibler
distance of Fourier spectra (see figure~\ref{power_mean_age}) or the distance
measure based on the mutual information displayed in
figure~\ref{mutual_mean_age}.  This normalization of the overall level of
synchronization is in parallel with the improvement of the psychiatric status.

\noindent The clustering behavior using mutual information
yields systematic lower values for patient $A$ --- at least in the first three
recordings --- which express a higher similarity between the signals of
the various channels compared to that of the control group. The results from
the fourth and the following recordings are comparable to those of the 
control patients. Figure~\ref{mutual_mean_age} shows the evolution of patient
$A$ under treatment using mutual information distance $D\left(M\right)$.\\
The discriminative power of a distance measure $D\left(Q\right)$ between two
groups ``0'' and ``1'' is expressed by  
\begin{equation}  
Contrast=\frac{\left(D^1\left(Q\right)-D^0\left(Q\right)\right)^2}
{\left(\sigma_D^1\left(Q\right)\right)^2+\left(\sigma_D^0\left(Q\right)\right)^2}
\label{equation_contrast}
\end{equation}
Table~\ref{result_table_overall} lists this contrast parameter between acute phase of
patient $A$ on one side (``0'') and control group $B$ (``1'') 
on the other side for the several derived measures.\\
The highest contrast in synchronization between these groups is achieved by Kullback-Leibler distances of the power spectra 
($Contrast=3.25$, see figure~\ref{power_mean_age}), followed by mutual information
($Contrast=1.42$, see figure~\ref{mutual_mean_age}). 
It is remarkable that Spearman's rank coefficients of correlation
($Contrast=0.64$) and Pearson's coefficients ($Contrast=0.16$) fail to discriminate between
the two groups. From the linear perspective the EEG-recordings of the
FLE-patient and the control group are undistinguishable and form in this sense
a homogenous group. 

\begin{center}
\begin{table}[tbp]{\bf \hspace{0.5cm} }\\[1ex]
\arrayrulewidth1.0pt
\renewcommand{\arraystretch}{1.3}
\begin{tabular}{|l|c|c|c||c|c|}
\hline
&  & \multicolumn{2}{|c||}{patient A} & \multicolumn{2}{c|}{control group $B$} \\ 
Method & Contrast & $D^A\left(Q\right)$ & $\sigma_D^A\left(Q\right)$ & $D^B\left(Q\right)$ & $\sigma_D^B\left(Q\right)$ \\ 
\hline
Kullback-Leibler       & 3.25 & 0.53 & 0.057 & 0.66 & 0.035 \\ \hline
Mutual Information     & 1.42 & 3.12 & 0.222 & 3.38 & 0.023 \\ \hline
Spearman Rank Order    & 0.64 & 1.43 & 0.012 & 1.42 & 0.004 \\ \hline
Pearson's coefficient  & 0.16 & 1.42 & 0.016 & 1.42 & 0.003 \\ \hline
\end{tabular}
\caption{\small\em Discrimination power between patient A
  (acute phase: first four EEG-recordings of patient $A$) and control group $B$ 
  expressed by the $Contrast$ (see Eq.~\protect\ref{equation_contrast}).
\label{result_table_overall} 
}
\end{table}
\end{center}

\subsection{Spatio-Temporal Characterization}
\begin{figure}[htb]
\epsfclipon
\epsfxsize=12.01cm
\epsfxsize=8.01cm
\epsfxsize=12.51cm
\centerline{\epsffile[0 0 446 169]{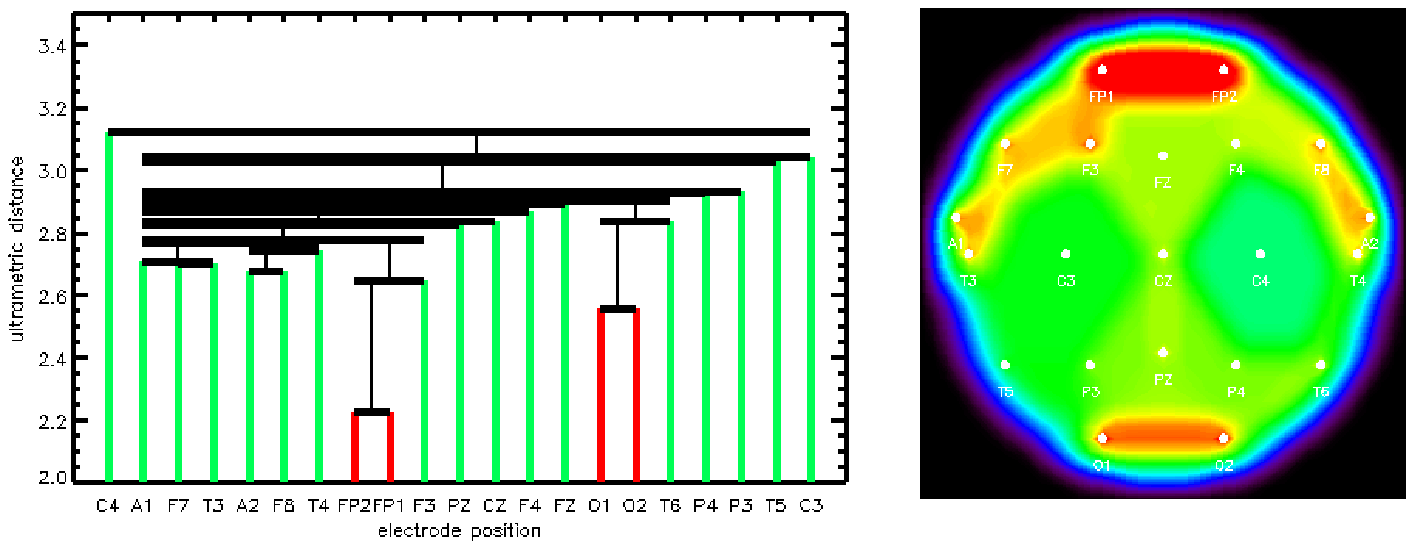}}
\caption{\small\em Spatio-temporal characteristic of brain activity based on
  mutual information of patient $A$ before seizure (see
  figure~\protect\ref{cor_plot_single_trace_ami}). This EEG recording correspond
  to A2 in figure~\protect\ref{topo_florian}. The left panel shows the exact 
  clustering scheme expressed by an hierarchical tree (labeled in red: FP1,
  FP2, O1, O2) while on the right side
  a two-dimensional approximation is presented.   
\label{patient_A_before_seizure} 
}
\end{figure}

\begin{figure}[htb]
\epsfclipon
\epsfxsize=12.01cm
\epsfxsize=8.01cm
\epsfxsize=12.51cm
\centerline{\epsffile[0 0 446 169]{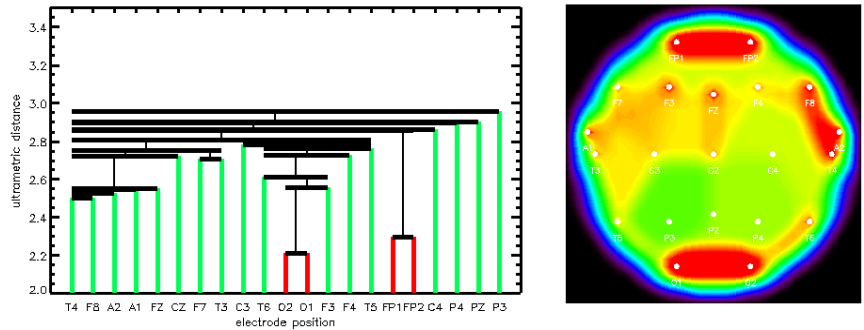}}
\caption{\small\em Spatio-temporal characteristic of brain activity based on mutual information of
  patient $A$ during seizure (see figure~\protect\ref{cor_plot_single_trace_ami}). This EEG recording 
correspond to A2 in figure~\protect\ref{topo_florian}. 
\label{patient_A_during_seizure} 
}
\end{figure}

\begin{figure}[htb]
\epsfclipon
\epsfxsize=12.01cm
\epsfxsize=8.01cm
\epsfxsize=12.51cm
\centerline{\epsffile[0 0 446 169]{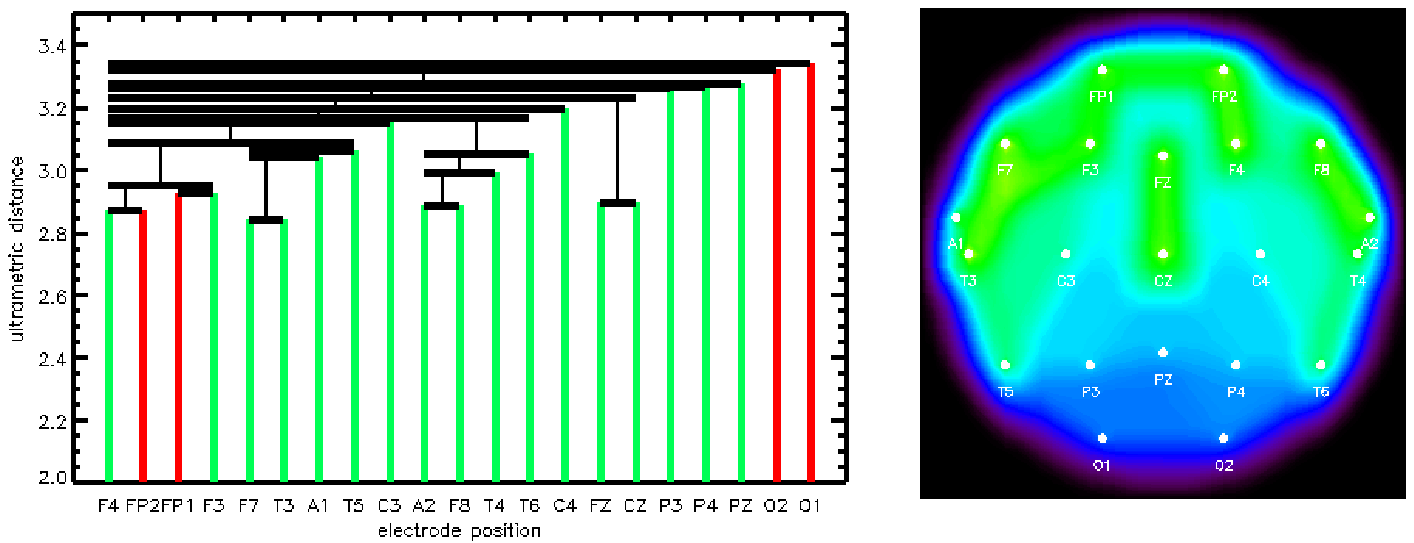}}
\caption{\small\em Representative example of spatio-temporal characteristic of brain
  activity based on mutual information from control group. This EEG recording correspond
  to B1 (sixth EEG, age 11.86 y) in figure~\protect\ref{topo_control}. In general,
  cluster depth decreases from the frontal to the occipital area.
\label{Control_group_B1_6} 
}
\end{figure}

The spatial structuring of the brain activity can be studied in more detail
by means of the hierarchical clustering as introduced in section~\ref{SectHierarch}. 
The hierarchical tree shows the relation between the signals obtained at the
several electrodes. Two states of patient $A$ (figure~\ref{patient_A_before_seizure}: before seizure,  
figure~\ref{patient_A_during_seizure}: during seizure) and a 
typical example of the control group (figure~\ref{Control_group_B1_6})
are shown. In the left panel the hierarchical tree is shown, while the
right panel gives a two dimensional approximation of the cluster behavior.
In general, an exact two-dimensional solution is not possible. The color coding of
these representations is identical in all figures: Cluster depth is color
coded by blue (loose cluster) over green, yellow to red (strong cluster).

\noindent The ultrametric distance of each branch can be
read from the ordinate. 
Following from top to bottom in the left panel, the clustering becomes
stronger, i.e. the items (time series) that belong to the same branch become
more similar to each other. Comparison of the figures show significant
differences in the spatial structuring. First of all, we
can see that the clusters are in general deeper in the FLE-patient than in the reference cases, as already shown in the
previous section. The deepest cluster, however, are formed by the signals
obtained in the frontal area (i.e. FP1-FP2) and in the occipital area (i.e.
O1 and O2). During seizure attack, the latter one is sometimes even the 
deepest cluster, as for instance in figure~\ref{patient_A_during_seizure}.
In contrast, in the control group the cluster depth decreases from the frontal
to the occipital area (see figure~\ref{Control_group_B1_6}). In
particular, the synchronization between the signals in O1 and O2 is rather poor in the reference cases, 
with respect to each other and with respect to the signals obtained at all other positions. Moreover, 
the aggregation level of O1 and O2 is very often at the highest ultrametric distance (see figure~\ref{Control_group_B1_6})

\noindent The typical representation in figure~\ref{Control_group_B1_6} shows that
normally --- and on average --- one observes a clusterization, which reflects the
geometrical placements quite well. The clusters are predominantly formed along
lines connecting the front side with the rear side, e.g.: FP1-F3, F7-T3-T5,
FZ-CZ. The synchronization between the hemispheres, which is clearly
visible in all cases, falls off backwards. In patient A, we observe stronger
clustering along horizontal lines, such as T6-O2-O1-T5, and a closer
clustering between geometrical more distant locations as for instance
between occipital-parietal and frontal areas.

\begin{figure}[htb]
\epsfclipon
\epsfxsize=12.01cm
\epsfxsize=8.01cm
\centerline{\epsffile[0 0 548 548]{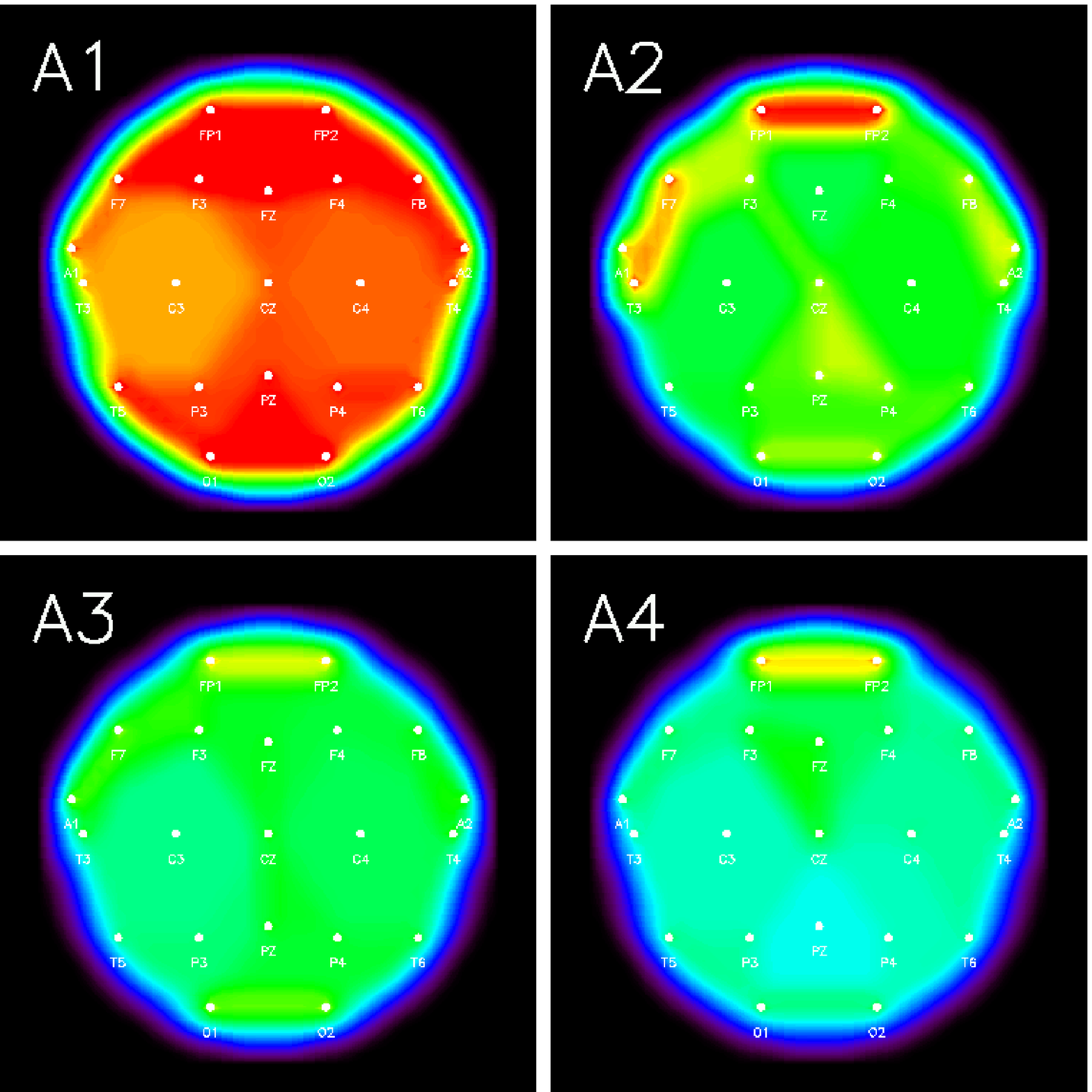}}
\caption{\small\em Change of mean spatio-temporal structuring of
  brain activity based on mutual information in patient $A$ reflecting
  the change of clinical features under therapy. A1 to A4 refers to the first
  four EEG of patient $A$ ordered by age (see table~\ref{patients_data}, 
  figure~\protect\ref{power_mean_age} or~\protect\ref{mutual_mean_age}): 
  A1 12.06 y, A2 12.10 y, A3 12.10 y, A4 12.16 y. The color coding is identical
  to that used in figures~\protect\ref{patient_A_before_seizure}--\protect\ref{Control_group_B1_6}.  
  Patient $A$ exhibit strong clustering along horizontal lines and close
  clustering between geometrical more distant locations as for instance
  between occipital-parietal and frontal areas. 
\label{topo_florian} 
}
\end{figure}

\begin{figure}[htb]
\epsfclipon
\epsfxsize=12.01cm
\epsfxsize=8.01cm
\centerline{\epsffile[0 0 548 548]{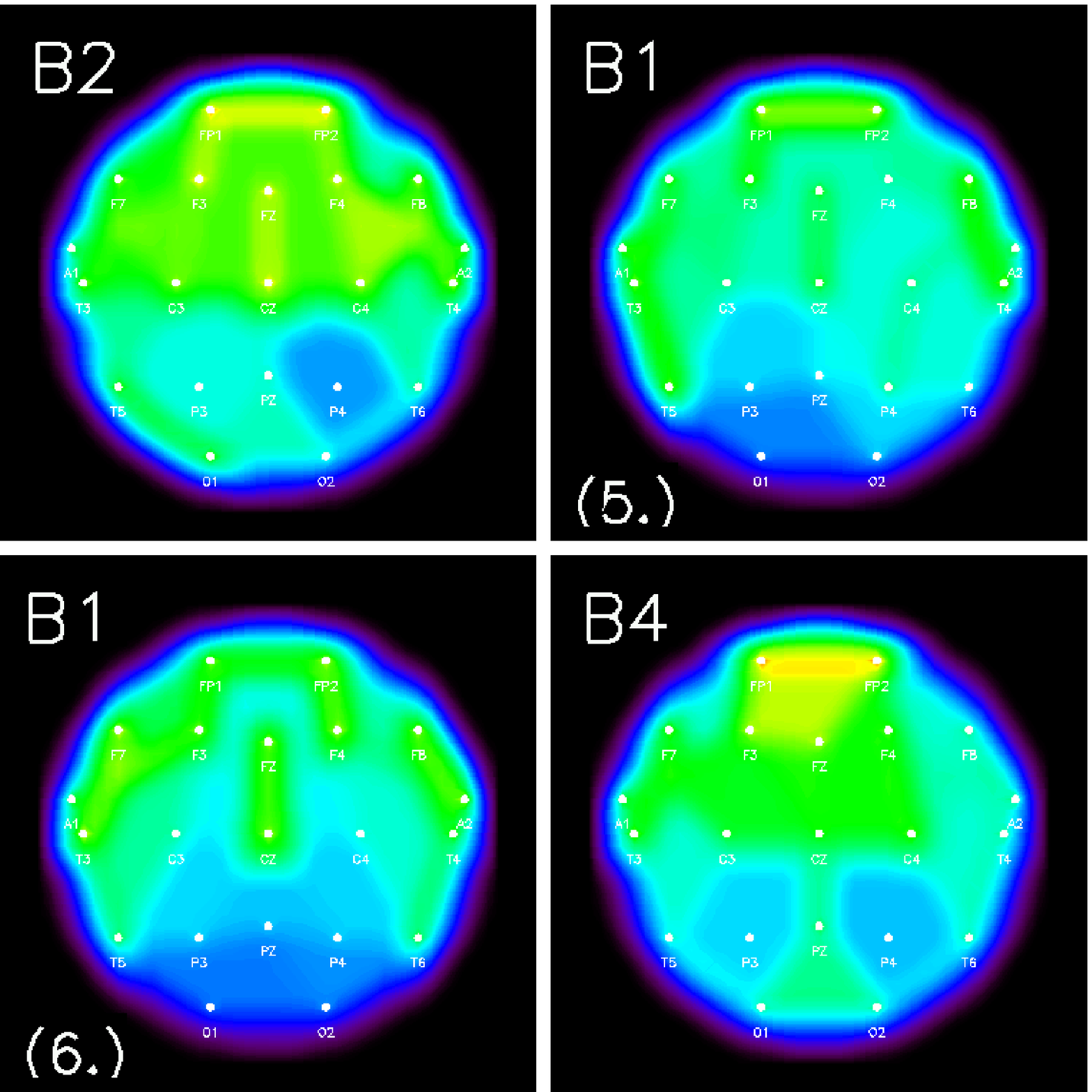}}
\caption{\small\em Representative examples of mean spatio-temporal structuring of
  brain activity based on mutual information in control group. B1 to B4 refers to the EEG of the
  control group ordered by age (see table~\ref{patients_data}, figure~\protect\ref{power_mean_age},
  ~\protect\ref{mutual_mean_age} and~\protect\ref{Control_group_B1_6}): 
  B2 11.28 y, B1 11.32 y (fifth EEG), B1 11.86 y (sixth EEG), B4 14.37 y. Brain activity in the
  control group clusters predominantly along the frontal-occipital
  direction with decreasing cluster strength. The color coding is identical
  to that used in figures~\protect\ref{patient_A_before_seizure}--\protect\ref{topo_florian}. 
\label{topo_control} 
}
\end{figure}

\noindent This is demonstrated by the two-dimensional visualization of the
cluster properties in the sequence of figure~\ref{topo_florian} (patient A) and
figure~\ref{topo_control} (control group). The images show the time-averaged cluster
behavior in succeeding EEG recordings: Thus these visualizations represent the mean
cluster properties of the corresponding EEG signals. The upper left image shows the
situation in patient $A$ before treatment. The overall degree of synchronization
is very high. After starting effective treatment with valproic acid and
lamotrigine clinical symptoms and seizures improved slowly. 
Only after readjustment of therapy with higher doses of valproic acid and lamotrigine the
cluster properties (lower panel) are getting similar to that of the control group. The lower right image is computed from an EEG recording after five
weeks of effective treatment. Even if there are still some accentuation in the
frontal area and in the occipital area, the two-dimensional
visualization is hardly distinguishable from that of the control group.

\section{Discussion}
The purpose of this investigation is the quantitative spatio-temporal 
characterization of brain activity from standard multi-channel scalp EEG recordings in a child 
suffering from frontal lobe epilepsy in contrast to a control group of age matched children 
without active epilepsy. From a diagnostic point of view an appropriate measure is 
missed, which quantitatively characterizes the status of the patient within the interictal 
phases. Therefore, the main objective was rather the description of the background 
activity, which is unsuspicious in usual visual inspection than the investigation 
of pre-seizure, seizure or post-seizure effects, their known accompanying electroencephalographic
signatures~\cite{detecting_epileptic_seizures_in_long-term_human_eeg_2008} or a prediction of seizures 
(see e.g. ~\cite{jerger_et_al0}, \cite{anticipation_of_epileptic_seizures_from_standard_EEG_recordings}, % 
\cite{seizure_anticipation_do_mathematical_measures_correlate_with_video-eeg_evaluation}).\\ 

\noindent The basic assumption of our approach is that the obvious increased 
synchronization level of the brain activity during attacks persists in a weakened form 
within interictal epochs, even if this is not easily seen by means of
conventional EEG analysis. 
Since the state of vigilance also affects strongly synchronization
of brain activity, e.g. sleep enhances synchronization, only periods with the
patients in a comparable state (i.e. awake, same assessment of level of attention) are included in the
analysis. To study epilepsy related synchronization in multichannel EEG recordings, 
we investigated the suitability of a number of different linear as well as
information based correlation measures (e.g. mutual information, Kullback-Leibler
distances). Spatio-temporal resolution of the derived measures is achieved in
the framework of the subsequent hierarchical cluster analysis.\\ 

\noindent In the case study of the patient suffering from frontal lobe epilepsy we find, 
that the synchronization level is significantly increased during its 
clinical manifestation with respect to all individuals of the control group.
Out of a number of methods in time and frequency domain, mutual information 
and Kullback-Leibler distances have the highest contrast between the control 
group and the interictal activity of the FLE patient. 
After successful medical suppression of the acute epileptic state the clinical picture still points to the 
presence of a frontal lobe syndrome. In consequence of a long term effective
therapy these symptoms disappeared. In the course of the two year lasting follow up of
the patient, the synchronization measures converge to the values
obtained for the members of the control group.\\  

\noindent In detail, just after starting the medical treatment we observe 
strong changes of the global measures with a clear tendency towards the range
defined by the reference group. Within the first weeks, however, the global and local measures still differ 
significantly from that of the control group, albeit the visual EEG inspection gives 
no hints to abnormal characteristics of background activity. After a short period of an about
two month lasting treatment the chosen measures are no more distinguishable from that
of the healthy probands in a statistical sense. Our results suggest that the
newly developed method could yield an appropriate parameter for the evaluation of the 
patients response to anticonvulsive drug treatment, which is independent of
classical epilepsy-related EEG features. From a clinical point of view, we found that 
the development of the derived parameters is in parallel with the process of the psychological 
normalization.\\ 

\noindent The interpretation of the spatial characteristics given by the detailed
cluster representation of the brain activity is more difficult. As far 
as the frontal lobe areas are concerned, our method confirms significantly an 
enhanced synchronization between the most relevant surface locations (i.e. FP1-FP2). Nevertheless, so
far our spatial reconstruction is not able to narrow down substantially the
assumed focal area. The most surprising aspect of our results, however, is
that we observe striking differences in the spatial structuring of the
occipital region (O1-O2), compared to the control group. This anomaly is clearly present in our 
characterizations, independent of the used measures, but beforehand undetected by the conventional 
EEG analysis.\\ 
A further result of our investigation is that in the pathological case of frontal lobe
epilepsy the synchronization levels along paths connecting the hemisphere are more pronounced with respect to the results of
the control group.
In the latter case we observe that the brain activity clusters along lines connecting frontal with rear (occipital)
parts. This phenomenon, in our opinion, resembles the so called secondary bilateral synchrony
(SBS), which is well known from literature~(\cite{tukel_jasper0}, \cite{blume_pillay0}).  SBS is frequently 
associated with a focus in the frontal lobes, but it is rather the result of a complex 
interaction of multiple foci as Blume and Pillay argue~\cite{blume_pillay0}. The presence of multiple foci in our
patient could not be settled.\\ 

\noindent In our opinion, the presented method can be helpful to extract additional 
diagnostic information out of inconspicuous background activity. The detailed analysis of EEG signals from
selected areas with the help of the shown cluster ansatz promises valuable and specific information on local 
synchronization features.\\
In general, we expect that the information gain of our approach benefits from a higher
spatial and temporal resolution of the data.\\

\noindent As a related application of our approach, we see the study of
neurodevelopmental disorders like autism. In this context Kulisek et al.~\cite{nonlinear_analysis_of_the_sleep_eeg_in_children_with_pervasive_developmental_disorder}
were able to characterize autistic children by  means of synchronization levels.\\

\newpage 
\leftline{\bf List of Figure Legends:}
\begin{enumerate}
\item {\small\em Time course of the mean distance $D\left(M_{t}\right)$ --- based on the
  mutual information --- for the second EEG recording of patient A
  (age$=12.10\ yr$). The according value of the averaged mean distance 
  $D\left(M_{t}\right)=3.103$ and standard deviation 
  $\sigma_D\left(M_{t}\right)=0.075$ (compare second observation of patient $A$ in
  figure~\protect\ref{mutual_mean_age}). At about $6.7\ min$ an epileptic 
  seizure (hatched area) starts. 
  The seizure attack is initiated by a loss of synchronization,
  while during the seizure the correlation is generally increased, thus the
  distance values are low. The red horizontal lines mark the mean value and standard
  deviation of the different EEG recordings of patient A. The blue horizontal
  lines label the corresponding quantities for the control group. Periods
  with interictal epileptiform discharges or ictal patterns, labeled in gray, are eliminated and do not enter
  the statistics.
} 
\item {\small\em Mean Kullback-Leibler distance for patient $A$ (acute phase: red,
  non-acute phase: green) and control group (blue) of the power spectra. The filled circles denote the averaged mean
  Kullback-Leibler distance $D\left(E\right)$, the error bars indicate the range 
  $\pm \sigma_D\left(E\right)$ of the single
  EEG recordings. The horizontal lines label the mean Kullback-Leibler
  distance for patient $A$ and control group, the dashed lines label the respective
  standard deviations.   
}
\item {\small\em Mean mutual Information for patient $A$ (acute phase: red,
  non-acute phase: green) and control group
  (blue). The filled circles denote the averaged mean mutual information
  $D\left(M\right)$, the error bars indicate the standard deviations $\sigma_D\left(M\right)$
  of the single EEG recordings. The horizontal lines label the mean mutual
  information for patient $A$ and control group, the dashed lines mark the
  respective standard deviations.   
}
\item {\small\em Spatio-temporal characteristic of brain activity based on
  mutual information of patient $A$ before seizure (see
  figure~\protect\ref{cor_plot_single_trace_ami}). This EEG recording correspond
  to A2 in figure~\protect\ref{topo_florian}. The left panel shows the exact 
  clustering scheme expressed by an hierarchical tree (labeled in red: FP1,
  FP2, O1, O2) while on the right side
  a two-dimensional approximation is presented.   
}
\item {\small\em Spatio-temporal characteristic of brain activity based on mutual information of
  patient $A$ during seizure (see figure~\protect\ref{cor_plot_single_trace_ami}). This EEG recording 
correspond to A2 in figure~\protect\ref{topo_florian}. 
}
\item {\small\em Representative example of spatio-temporal characteristic of brain
  activity based on mutual information from control group. This EEG recording correspond
  to B1 (sixth EEG, age 11.86 y) in figure~\protect\ref{topo_control}. In general,
  cluster depth decreases from the frontal to the occipital area.
}
\item {\small\em Change of mean spatio-temporal structuring of
  brain activity based on mutual information in patient $A$ reflecting
  the change of clinical features under therapy. A1 to A4 refers to the first
  four EEG of patient $A$ ordered by age (see table~\ref{patients_data}, 
  figure~\protect\ref{power_mean_age} or~\protect\ref{mutual_mean_age}): 
  A1 12.06 y, A2 12.10 y, A3 12.10 y, A4 12.16 y. The color coding is identical
  to that used in figures~\protect\ref{patient_A_before_seizure}--\protect\ref{Control_group_B1_6}.  
  Patient $A$ exhibit strong clustering along horizontal lines and close
  clustering between geometrical more distant locations as for instance
  between occipital-parietal and frontal areas. 
}
\item {\small\em Representative examples of mean spatio-temporal structuring of
  brain activity based on mutual information in control group. B1 to B4 refers to the EEG of the
  control group ordered by age (see table~\ref{patients_data}, figure~\protect\ref{power_mean_age},
  ~\protect\ref{mutual_mean_age} and~\protect\ref{Control_group_B1_6}): 
  B2 11.28 y, B1 11.32 y (fifth EEG), B1 11.86 y (sixth EEG), B4 14.37 y. Brain activity in the
  control group clusters predominantly along the frontal-occipital
  direction with decreasing cluster strength. The color coding is identical
  to that used in figures~\protect\ref{patient_A_before_seizure}--\protect\ref{topo_florian}. 
}
\end{enumerate}
\end{document}